# Towards a Blockchain-Based CI/CD Framework to Enhance Security in Cloud Environments

Sabbir M. Saleh[1][a], Nazim Madhavji[1][b] and John Steinbacher[2][c]
*[1]Department of Computer Science, University of Western Ontario, London, Ontario, Canada*
*[2]IBM Canada Lab, Markham, Ontario, Canada*
ssaleh47@uwo.ca, madhavji@gmail.com, jstein@ca.ibm.com

Abstract: Security is becoming a pivotal point in cloud platforms. Several divisions, such as business organisations, health care, government, etc., have experienced cyber-attacks on their infrastructures. This research focuses on security issues within Continuous Integration and Deployment (CI/CD) pipelines in a cloud platform as a reaction to recent cyber breaches. This research proposes a blockchain-based solution to enhance CI/CD pipeline security. This research aims to develop a framework that leverages blockchain's distributed ledger technology and tamper-resistant features to improve CI/CD pipeline security. The goal is to emphasise secure software deployment by integrating threat modelling frameworks and adherence to coding standards. It also aims to employ tools to automate security testing to detect publicly disclosed vulnerabilities and flaws, such as an outdated version of Java Spring Framework, a JavaScript library from an unverified source, or a database library that allows SQL injection attacks in the deployed software through the framework.

## 1 INTRODUCTION

Cloud computing is crucial for software development because of its scalability and efficiency. Its continuous interactions between service providers and users make it vital for modern software engineering.

Continuous Integration and Deployment (CI/CD) pipelines are central to this ecosystem. These automate the concept of Continuous Software Engineering for building, testing, and deploying (Fitzgerald & Stol, 2017).

However, the increasing reliance on these pipelines has exposed significant security vulnerabilities (Saleh et al., 2024), as evidenced by breaches in tools such as GitHub Actions, SonarQube, Harbor Registry, and Docker Containers.

Previous studies have shown that CI/CD pipeline tools have vulnerabilities, as evidenced by image manipulation (80% of Docker Hub images reveal high-level vulnerabilities discovered through scanning more than 300,000 images in around 85,000 repositories) (Shu et al., 2017), unauthorised access and lack of authentication in the Harbor Registry (Mahboob & Coffman, 2021) to both public and private image repositories, posing a security risk.

Such vulnerabilities make CI/CD pipelines attractive targets for cyberattacks, including supply chain attacks, ransomware (Saboor et al., 2022), denial-of-service exploits, etc. (Drees et al., 2021).

Prominent examples such as the Log4j exploit, SolarWinds breach, and CodeCov incident (Williams, 2022; Benedetti et al., 2022) highlight the urgent need for robust security mechanisms in CI/CD pipelines.

This research proposes a blockchain-based framework integrating Hyperledger Fabric with Jenkins to enhance CI/CD pipeline security. By leveraging the blockchain's distributed ledger and tamper-resistant properties, the framework aims to ensure artefact immutability, restrict unauthorised access, and automate vulnerability detection.

Key features include detecting outdated or unverified dependencies, preventing SQL injection attacks, and securing the deployment process against unauthorised modifications.

[a] https://orcid.org/0000-0001-9944-2615
[b] https://orcid.org/0009-0006-5207-3203
[c] https://orcid.org/0009-0001-6572-6326

The proposed methodology involves developing a prototype blockchain solution tailored for CI/CD pipelines, conducting experiments with datasets containing insecure code and mock attacks, and assessing the framework's impact on security performance.

Initial results demonstrate the framework's feasibility, emphasising its potential to address critical gaps in secure software deployment.

This study also explores the broader implications of integrating Hyperledger Fabric into CI/CD workflows, focusing on usability, developer adoption, and scalability.

The framework's permissioned blockchain approach ensures that only authorised users can access sensitive data, further enhancing the pipeline's security.

The research aims to answer the key question: "*How integrating Hyperledger Fabric with CI/CD does pipelines impact security performance?*"

The rest of the paper is as follows:-

Section 2 reviews blockchain-integrated DevOps and CI/CD security and identifies key gaps. Section 3 outlines challenges in blockchain adoption for CI/CD and plans a security framework using Hyperledger Fabric.

Section 4 details the research methodology, including prototype development, threat modelling, and security assessments. Sections 5 and 6 present the preliminary results of integrating Hyperledger Fabric with Jenkins, highlighting security improvements and challenges.

Section 7 concludes with contributions and future work with AI-driven anomaly detection, expanded testing, and real-time vulnerability scanning.

## 2 RELATED WORK

While working on this research, we acknowledged some works where reviews reported on or proposed frameworks for blockchain-integrated DevOps and CI/CD pipelines.

A literature review (Akbar et al., 2022) explored the potential benefits of blockchain-adopted DevOps technology, such as decentralisation, automation of smart contracts, faster CD, etc. The study also proposed a framework that has yet to focus on the security aspects of the CI/CD pipeline over the cloud.

Bankar and Shah (2021) developed a framework for integrating blockchain into the DevOps process to make the software industry's workplace more responsible and reasonable. However, it has yet to explore CI/CD in cloud environment pipelines to enhance security.

Farooq and Usman (2023) proposed a blockchain-based framework for DevOps, using technologies such as IPFS (InterPlanetary File System), smart contracts, and Jenkins. This framework may help distributed or remote teams securely communicate and share data and can improve the development and operation process; however, it has yet to enhance the security of cloud platforms.

Wohrer and Zdun (2021) investigated blockchain-oriented DevOps's current usage and methods by combining grey literature and DevOps studies. While giving insights into the Ethereum (smart contract platform) and the Solidity programming language, this study limits the CI/CD pipeline security issues, such as unauthorised access, data breaches, and vulnerabilities over cloud platforms.

While various works and frameworks exist to integrate blockchain technology with DevOps and CI/CD, as mentioned in Saleh et al., 2024, there are still gaps (e.g., advanced security mechanisms for containerised applications, low-code platforms, GitHub Actions, etc.) and challenges (e.g., weak authentication, dependency risks, security tool integration, etc.) in addressing security issues on cloud platforms.

## 3 RESEARCH GAPS AND OBJECTIVES

CI/CD tools, such as Jenkins, GitHub Actions, and GitLab CI/CD, lack built-in support for blockchain technologies. This leads to integration complexities such as manual configurations, native plugins, and limited interoperability with existing workflows.

The high costs associated with blockchain integration—transaction fees, resource-intensive deployments, and custom development—pose barriers to scalability and broader industry adoption.

Our research provides valuable insights into these dynamics and addresses these gaps by developing a blockchain-based prototype that integrates Hyperledger Fabric with the Jenkins CI/CD pipeline.

This approach emphasises secure artefact management, tamper-proof build processes, and automated vulnerability detection, underscoring its potential to transform CI/CD pipeline security. By demonstrating blockchain integration's feasibility and practical impact, this work lays the foundation for broader adoption and further innovation in this area.

The table below highlights key blockchain integration-related work with DevOps and CI/CD pipelines. These studies provide valuable insights into leveraging blockchain to enhance transparency, collaboration, and reliability in software development.

However, they also reveal significant research gaps, particularly regarding the security enhancement of CI/CD pipelines and the integration of specific tools such as Hyperledger Fabric and Jenkins or Tekton. Table 1 outlines the relation of these works to the current research and identifies the unexplored areas.

Table 1: Relationship of Existing Works to Current Research

| Work | Relation to this work | Research Gap |
| --- | --- | --- |
| Toward Effective and Efficient DevOps using Blockchain (Akbar et al., 2022): This paper highlights blockchain benefits in decentralisation, automation, and faster delivery by proposing a framework leveraging smart contracts for transparency and reliability. | Provides insights into potential frameworks for integrating blockchain with DevOps, aligning with CI/CD security research. | It focuses on DevOps processes but lacks specific emphasis on CI/CD pipeline security. |
| Blockchain-Based Framework for Software Development Using DevOps (Bankar & Shah, 2021): Integrates Hyperledger Composer and Fabric into DevOps with tools such as IPFS, enabling traceability, transparency, and secure file management. | Showcases blockchain integration benefits in DevOps, offering insights for CI/CD pipeline security enhancement. | The comprehensive DevOps framework has not yet addressed security enhancement in CI/CD pipelines. |
| Harnessing the Potential of Blockchain in DevOps (Farooq & Usman, 2023): Proposes a Distributed DevOps framework using IPFS, smart contracts, and Jenkins for secure and transparent development in distributed teams. | Aligns with CI/CD research by combining blockchain and DevOps to enhance collaboration and transparency. | It focuses on distributed DevOps but lacks integration with Hyperledger Fabric, Jenkins, or Tekton. |
| DevOps for Ethereum Blockchain Smart Contracts (Wohrer & Zdun, 2021): Explores testing, CI/CD practices, and deployment for smart contracts, emphasising automated pipelines and monitoring tools like blockchain explorers. | Offers insights for CI/CD practices and blockchain applications relevant to Hyperledger Fabric integration research. | Focuses on Ethereum-based smart contracts, not Hyperledger Fabric or CI/CD security integration. |

## 4 RESEARCH METHODOLOGY

The core goal is to see if blockchain integration with the CI/CD pipeline can improve security in cloud environments by fortifying the pipeline.

### 4.1 Research Approach

To better understand the shortcomings of existing security practices, we reviewed an anonymous number of literature. We reviewed literature (Saleh et al., 2024) on security difficulties in CI/CD pipelines, such as supply chain attacks Log4j, SolarWinds, xz utils incidents, etc. (Williams et al., 2024).

We also reviewed the literature (Saleh et al., 2024) on existing approaches to securing the CI/CD pipeline by integrating blockchain.

The reviewed approaches for securing the CI/CD pipeline through blockchain integration include leveraging modular architectures. Hyperledger Fabric is used for scalability and security, Hyperledger Caliper for performance benchmarking, and automated workflows using Jenkins and Kubernetes.

Private data collections, encryption, and point-to-point transactions enhance privacy, while Truffle Suite and OpenZeppelin enable secure smart contract development.

Vulnerability testing tools such as FUDGE and FuzzGen ensure secure validation, and Algorand focuses on resource efficiency.

These approaches collectively address scalability, security, and efficiency challenges in blockchain-integrated CI/CD environments. From there, we shall develop our prototype and achieve research results.

### 4.2 Research Plan

Our primary focus is experimenting with integrating blockchain and CI/CD pipeline using toy examples of code and mock attacks (such as attempts to tamper with code, unauthorised access, etc.). The idea is to obtain preliminary feedback and assess the degree.

This way, we can assess the blockchain's immutability and consensus mechanism and create blocks in the pipeline to prevent such attacks from being ingrained or spread.

Hyperledger Fabric is best suited for CI/CD pipelines in terms of privacy, scalability, integration, security, and cost efficiency.

Table 2 compares the features of blockchain platforms such as Hyperledger Fabric, Ethereum, Quorum, etc. based on previous studies (Mohammed et al., 2021; Dabbagh et al., 2020; Ucbas et al., 2023; Polge et al., 2021; Singh et al., 2023; Valenta & Sandner, 2017; Dar et al., 2023)

Table 2: Framework Comparison for CI/CD Pipelines

| Feature | Hyperledger Fabric | Ethereum | Quorum | R3 Corda | Tetherum |
|---|---|---|---|---|---|
| Security | Permissioned blockchain with robust identity management (TLS, certificates). | Public, decentralised with limited default security for private enterprise needs. | Enhanced security for private networks with permissioned access. | High security through restricted transactions and granular privacy controls. | Focused on security with AI-driven monitoring and anomaly detection capabilities. |
| Privacy | Private data collections, channel-based transactions. | Public blockchain with limited privacy options. | Supports private transactions for confidentiality. | Point-to-point transactions, restricted data sharing. | Data isolation and permissioned subnets for high privacy. |
| Scalability | Highly scalable for enterprise environments with modular architecture. | Scalability challenges due to global consensus and high resource consumption. | Scales well in private networks with reduced consensus complexity. | Effective for small, targeted networks yet unsuitable for open systems. | Scales efficiently for real-time, high-performance workflows. |
| Integration | Seamless integration with CI/CD tools (e.g., Jenkins, Kubernetes, Docker). | It integrates with CI/CD tools, but high costs and scalability issues limit its adoption. | Strong for privacy-focused CI/CD pipelines in enterprise environments. | Suitable for financial/legal contract validation workflows in CI/CD pipelines. | Platform with promising CI/CD integration for data-intensive, AI-driven pipelines. |
| Cost Efficiency | Cost-effective due to its permissioned nature. | High transaction costs and resource consumption. | Lower costs compared to Ethereum, designed for private networks. | Cost-efficient for specific use cases but not in general. | High efficiency in terms of cost and resources for data-heavy applications. |

Then, we shall experiment with threat modelling frameworks such as STRIDE, PASTA, LINDDUN, etc., to determine the pipeline's robustness.

These frameworks can be applied individually or in combination (e.g., STRIDE + PASTA, LINDDUN + PASTA, PASTA + OCTAVE, etc.) to assess the security of CI/CD pipelines, particularly in scenarios involving sensitive data, third-party integrations, and complex deployment workflows.

We shall also apply coding standards (secure coding practices) to prevent attacks such as SQL injection, Cross-Site Scripting (XSS), Hardcoded Credentials, etc.

Table 3 compares the widely used threat modelling frameworks and their relevance to CI/CD pipelines.

Table 3: Framework Comparison for CI/CD Pipelines

| Framework | Key Strengths | Use in CI/CD |
|---|---|---|
| STRIDE | Comprehensive threat categorisation | Identifies threats at various pipeline stages |
| PASTA | Risk and business-impact focus | Simulates attacks on deployment systems |
| LINDDUN | Privacy-centric | Secures sensitive data in CI/CD workflows |
| OCTAVE | Organisational and operational focus | Identifies systemic pipeline risks |
| Attack Trees | Visual representation of attack paths | Models potential pipeline vulnerabilities |
| DevOps-Specific Models | Tailored to CI/CD pipelines | Secures all stages from version control to deployment |

## 5 DEVELOPMENT AND IMPLEMENTATION OVERVIEW

This section details integrating Hyperledger Fabric with the Jenkins CI/CD pipeline. The process incorporates requirements allocation, designing, implementation, testing and evaluation. These efforts collectively form the foundation of our prototype.

### 5.1 System Development

Our contribution interfaces with the system by serving as a security layer within the CI/CD pipeline, ensuring the integrity and privacy of software artefacts throughout the development process.

The contextual diagram (Figure 1) visually represents our system within its intended environment. It illustrates the integration of Hyperledger Fabric with Jenkins as part of the broader CI/CD pipeline.

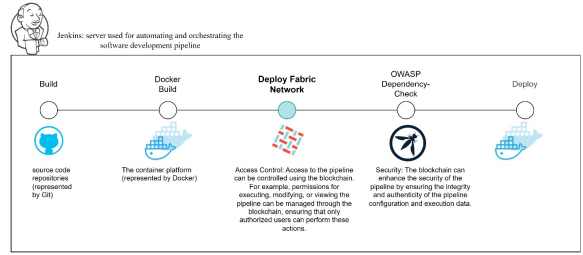

Figure 1: Contextual Diagram of the System

The class diagram (Figure 2) represents the system's structural overview, depicting the key components, attributes, and relationships.

We set up the development environment using Docker, Git, Jenkins, and Hyperledger Fabric as the project's foundation, created cryptographic materials, configured peers and orderers, and the necessary channels.

Next, we developed smart contracts (chaincode) tailored to our use case. Following that, we integrated Jenkins with Hyperledger Fabric, setting up Jenkins jobs to interact with the blockchain, covering tasks such as installing, initiating, invoking chaincode functions, and querying the ledger.

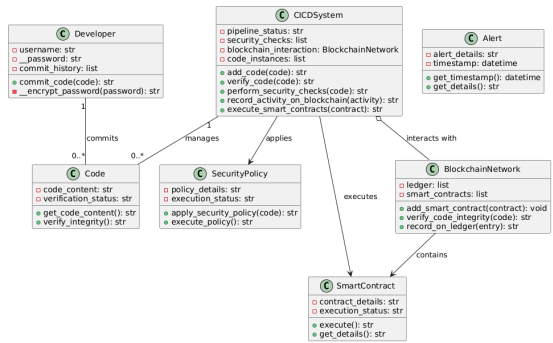

Figure 2: Class Diagram of the System

We containerised the application and its services using Docker by writing Dockerfiles, creating docker-compose configurations, building Docker images, and running the required containers to ensure smooth system operation.

We integrate OWASP Dependency Check as part of our security measures to identify vulnerabilities in third-party dependencies within our pipeline.

Integrating Dependency Check into the CI/CD pipeline allows us to continuously scan libraries and dependencies from the National Vulnerability Database (NVD) for known security issues.

This helps us proactively address vulnerabilities and reduce the risk of introducing insecure components into our system. This makes the prototype resilient and protected against potential threats, such as remote code execution, data breaches, or privilege escalation, from outdated or vulnerable external libraries, such as Log4j, Struts 2, or Jackson.

We obtained an NVD API Key to access the National Vulnerability Database (NVD) directly. This API key allows us to integrate more seamlessly with the NVD and retrieve the most up-to-date information on vulnerabilities associated with our dependencies.

The flowchart (Figure 3) illustrates the steps of our Jenkins pipeline. It showcases how each stage (build, security, deploy) contributes to the overall system.

## 5.2 Pipeline Overview and Stages

The Jenkins pipeline consists of five distinct stages, each playing a crucial role in the CI/CD process:

Build Stage: Compiles the code, transforming the test source file into an executable. This foundational step ensures the application code is built correctly before proceeding.

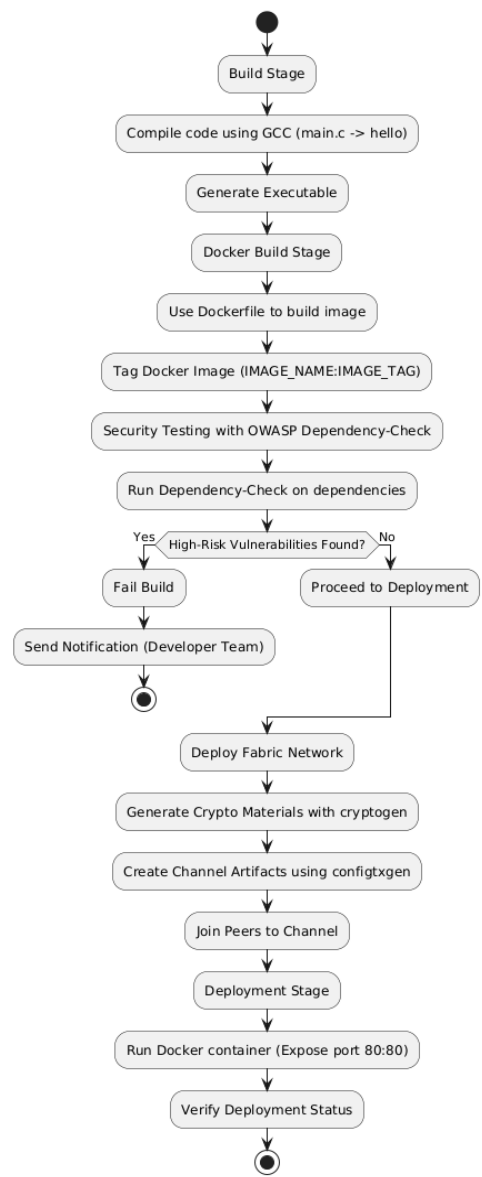

Figure 3: Jenkins Pipeline Flowchart

Docker Build Stage: Containerises the compiled application. The Docker image was built using the Dockerfile, isolating dependencies to confirm that the system runs consistently across different environments.

OWASP Dependency-Check Stage: Runs OWASP Dependency-Check to scan project dependencies for vulnerabilities. If high-risk issues are detected (e.g., SQL Injection, XSS, Insecure

Deserialization), the pipeline halts, preventing insecure deployments.

Deploy Fabric Network Stage: This stage automates the setup of the Hyperledger Fabric by generating cryptographic materials using cryptogenic, which creates the necessary cryptographic assets for secure communication.

We also configured channels using configtxgen and joined peers to them, streamlining the traditionally manual setup process and ensuring consistency across the blockchain.

Deployment Stage: This stage deploys the containerised application and makes the application live for further integration and testing.

### 5.3 Pipeline Settings and Modularity

Environment Variables:

FABRIC_BIN: Directory path for the blockchain binaries.

IMAGE_NAME and IMAGE_TAG: Manage Docker image name and version, ensuring consistency.

CONTAINER_NAME: Dynamically names containers using build specifics, which is helpful for version tracking and parallel deployments.

Using '*Agent Any*' allows the pipeline to run on any Jenkins agent, providing flexibility and scalability—ideal for adapting to infrastructure changes or using cloud-based agents. Figure 4 illustrates the pipeline's visual representation.

## 6 RESULTS, DISCUSSION, AND ANALYSIS

The results from the initial integration of Hyperledger Fabric with Jenkins demonstrate the promising capabilities for enhancing CI/CD pipeline security.

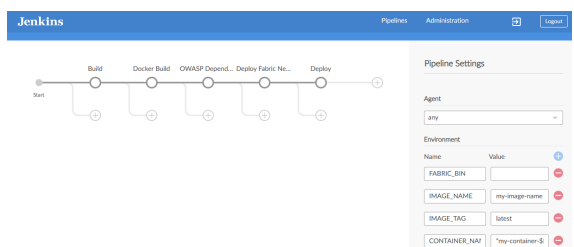

Figure 4: The Jenkins Pipeline

The following diagram (Figure 5) is vital to understanding the performance impact of integrating security processes into the CI/CD pipeline.

Security Metrics: The Dependency Check stage adds time to the overall pipeline, but this is an acceptable trade-off considering the added security.

Average Stage Times: This duration for the dependency check indicates the depth of the analysis, which is crucial for identifying potential vulnerabilities in third-party libraries.

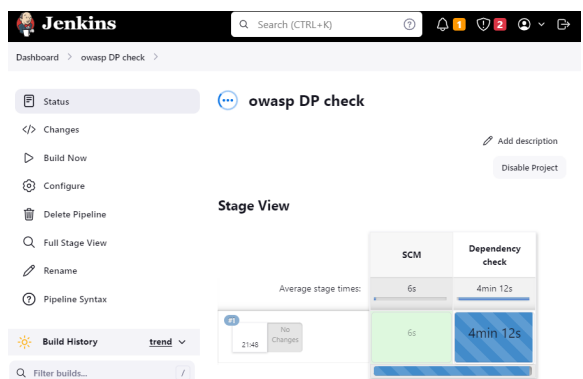

Figure 5: Initial Testing Output of the OWASP Dependency Check

While the delay can be seen as unfavourable, it is a minor cost compared to the security gains in detecting and mitigating potential threats in dependencies, such as an outdated version of Java Spring Framework, a JavaScript library from an unverified source, or a database library that allows SQL injection attacks in the deployed software through the pipeline.

The complexity of integrating Hyperledger Fabric with Jenkins may introduce issues, such as synchronisation problems, dependency conflicts, or network setup challenges. These issues can also be related to implementation errors, such as incorrect chaincode deployment, cryptographic material mismanagement, or inefficient peer ordering.

Configuration challenges, such as mismatched environment variables, misaligned channel policies, or improper Jenkins job triggers, may also exist.

### 6.1 Implications of the Research Results

The prototype offers significant implications for secure software development in academia and industry intending to tackle vulnerabilities.

Current standard CI/CD processes may not effectively mitigate issues such as unauthorised code changes and tampering, as numerous attacks occur regularly.

This research may lead to the widespread adoption of blockchain technologies in CI/CD practices, thereby enhancing software security and privacy across industries that rely heavily on the cloud.

## 6.2 Analysis

The OWASP Dependency Check stage was tested in the pipeline to ensure the dependencies were secure and verified against common vulnerabilities.

*Performance Impact:*

- Running the OWASP Dependency Check takes some minutes, which adds time to the overall deployment process.
- In a situation where deployment needs to be quick, this added time could be an issue.

*Improving Efficiency:*

To reduce this impact, we plan to make the process faster by:

- Running tasks in parallel (multiple things simultaneously).
- Using more efficient security tools (Snyk or Trivy) that might take less time.

*Usability Challenges:*

Setting up Hyperledger Fabric with Jenkins is not straightforward. It requires a lot of customisation and effort to get them working together properly.

Despite these difficulties, once it is set up, the added security benefits of utilising blockchain (transparency and preventing tampering) are precious, especially for clouds that need a high level of security.

## 7 CONCLUSIONS AND FUTURE WORKS

This research aims to enhance security in CI/CD pipelines through blockchain integration. Our findings demonstrate that integrating Hyperledger Fabric with Jenkins improves the detection and mitigation of security vulnerabilities in software development processes. These can collectively enhance the security of the cloud.

The future works for this research focus on the following:-

We shall refine the prototype by adding more security features, such as integrating AI-based anomaly detection techniques from our previous work (Saleh et al., 2024) to improve data privacy. A future technical paper details the integration process, including challenges, architecture, and strategies.

We plan to expand the testing process by using diverse code from different programming languages and frameworks, leveraging AI models to analyse and detect anomalies in insecure code and real-world datasets.

Mock attacks, such as ReDoS and supply chain attacks, shall be tested for system resilience. Integration of OWASP ZAP (Zed Attack Proxy) shall automate security testing of smart contracts, web applications, and APIs.

Furthermore, integrating CVE (Common Vulnerabilities and Exposures)/CPE (Common Platform Enumeration) APIs shall enable practical vulnerability scanning to ensure a robust security framework for CI/CD pipelines.

## ACKNOWLEDGEMENTS

We acknowledge Isra'a Al-Abbasi, an undergraduate student at Western University, for her dedicated efforts in implementing and assisting with the technical aspects of this work. Isra'a's work executing and refining these implementations was crucial to advancing the research, and their commitment to the work was influential in bringing the research to an accomplishment.